# Title: On-the-fly Closed-loop Autonomous Materials Discovery via Bayesian Active Learning

**Authors:** A. Gilad Kusne[‡,1,2,*], Heshan Yu[‡,2], Changming Wu[3], Huairuo Zhang[4,5], Jason Hattrick-Simpers[1], Brian DeCost[1], Suchismita Sarker[6], Corey Oses[7], Cormac Toher[7], Stefano Curtarolo[7], Albert V. Davydov[4], Ritesh Agarwal[8], Leonid A. Bendersky[4,5], Mo Li[3], Apurva Mehta[6], Ichiro Takeuchi[2,9,*]

**Affiliations:**

[1] Materials Measurement Science Division, National Institute of Standards and Technology, Gaithersburg, MD 20899, US

[2] Materials Science and Engineering Department, University of Maryland, College Park, MD 20742, US

[3] Electrical & Computer Engineering Department, University of Washington, Seattle, WA 98195, US

[4] Materials Science and Engineering Division, National Institute of Standards and Technology, Gaithersburg, MD 20899, US

[5] Theiss Research, Inc., La Jolla, CA 92037, US

[6] Stanford Synchrotron Radiation Lightsource, SLAC National Accelerator Laboratory, Menlo Park, CA 94025, US

[7] Mechanical Engineering and Materials Science Department and Center for Autonomous Materials Design, Duke University, Durham, NC 27708, US

[8] Materials Science and Engineering Department, University of Pennsylvania, Philadelphia, PA 19104, US

[9] Maryland Quantum Materials Center, University of Maryland, College Park, MD 20742, US

* Emails: aaron.kusne@nist.gov; takeuchi@umd.edu

‡ These authors contributed equally to this work.

**Abstract:** Active learning – the field of machine learning (ML) dedicated to optimal experiment design, has played a part in science as far back as the 18th century when Laplace used it to guide his discovery of celestial mechanics[1]. In this work we focus a closed-loop, active learning-driven autonomous system on another major challenge, the discovery of advanced materials against the exceedingly complex synthesis-processes-structure-property landscape. We demonstrate autonomous research methodology (i.e. autonomous hypothesis definition and evaluation) that can place complex, advanced materials in reach, allowing scientists to fail smarter, learn faster, and spend less resources in their studies, while simultaneously improving trust in scientific results and machine learning tools. Additionally, this robot science enables science-over-the-network, reducing the economic impact of scientists being physically separated from their labs. We used the real-time closed-loop, autonomous system for materials exploration and optimization (CAMEO) at the synchrotron beamline to accelerate the fundamentally interconnected tasks of rapid phase mapping and property optimization, with each cycle taking seconds to minutes, resulting in the discovery of a novel epitaxial nanocomposite phase-change memory material.

**Main Text:**
Technologies drive the perpetual search for novel and improved functional materials, necessitating the exploration of increasingly complex multi-component materials[2]. With each new component or materials parameter, the space of candidate experiments grows exponentially. For example, if investigating the impact of a new parameter (e.g. introducing doping) involves approximately ten experiments over the parameter range, $N$ parameters will require on the order of $10^{N+}$ possible experiments. High-throughput synthesis and characterization techniques offer a partial solution: with each new parameter, the number of candidate experiments rapidly escapes the feasibility of exhaustive exploration. The search is further confounded by the diversity and complexity of materials composition-structure-property (CSP) relationships including materials-processing parameters and atomic disorder[3]. Coupled with the sparsity of optimal materials, these challenges threaten to impede innovation and industrial advancement.

Structural phase maps, which describe the dependence of materials structure on composition, serve as blueprints in the design of functional and structural materials, as most materials properties are tied to crystal-structure prototypes. For example, property extrema tend to occur within specific phase regions (e.g. magnetism and superconductivity) or along phase boundaries (e.g. caloric-cooling materials and morphotropic phase-boundary piezoelectrics). Structural phase maps, and more specifically equilibrium phase diagrams, were traditionally generated over years with point-by-point Edisonian approaches involving iterative materials synthesis, diffraction-based structure characterization, and crystallographic refinement.

Machine learning (ML) is transforming materials research before our eyes[4], and yet direct coupling of ML with experiments remains a formidable challenge. Closed-loop Autonomous System for Materials Exploration and Optimization (CAMEO) offers a new materials research paradigm to truly harness the accelerating potential of ML, setting the stage for the 21$^{st}$-century paradigm of materials research – the autonomous materials research lab run under the supervision of a robot scientist or artificial scientist[5].

Active learning[6] – the ML field dedicated to optimal experiment design (i.e. adaptive design), is key to this new paradigm. Active learning provides a systematic approach to identify the best experiments to perform next to achieve user-defined objectives. Bayesian optimization (BO) active learning techniques have been used more recently to guide experimentalists in the lab to optimize unknown functions[7–12]. BO methods balance the use of experiments to explore the unknown function with experiments that exploit prior knowledge to identify extrema. However, these past studies only advised researchers on the next experiment to perform, leaving experiment planning, execution, and analysis to the researcher. More recently, autonomous systems have been demonstrated for process optimization[13, 14] and sample characterization[15]. Taking another step and placing active learning in real-time control of solid-state materials exploration labs promises to accelerate materials discovery while also rapidly and efficiently illuminating complex materials-property relationships.

We demonstrate CAMEO in real-time control of X-ray diffraction measurement experiments over composition spreads at the synchrotron beamline and in the lab. The algorithm accelerates phase mapping and materials discovery of a novel solid-state material, with a 10-fold reduction in required experiments, each iteration taking tens of seconds to tens of minutes depending on the experimental task. It uses an active-learning campaign that exploits physics knowledge (e.g. Gibbs phase rule) and combines the joint objectives of maximizing knowledge of the phase map $P(x)$ with hunting for materials $x_*$ that correspond to property $F(x)$ extrema, while also exploiting the shared information across these tasks via function $g$ (See Eq 1). Here $x \in \mathbb{R}^d$ is the set of $d$ materials-composition parameters. This allows CAMEO to target its search in specific phase regions or to search near phase boundaries, thus exploiting the dependence of materials property on structure. We demonstrate that this physics-informed approach accelerates materials optimization compared to general optimization methodologies that focus on directly charting the high dimensional, complex property function.

$$x_* = \mathrm{argmax}_x \left[ g(F(x), P(x)) \right] \quad Eqn\ (1)$$

Here, we explored the Ge-Sb-Te ternary system to identify an optimal phase-change memory (PCM) material for photonic switching devices[16]. PCM materials can be switched between the amorphous and crystalline states with an associated change in resistance and optical contrast which can be accessed on the nanosecond scale or shorter. Various Ge-Sb-Te based PCMs, especially $Ge_2Sb_2Te_5$ (GST225), have been used in DVD-RAM and nonvolatile phase-change random-access memory. We have implemented our strategy for identifying the optimal composition within the ternary for high-performance photonic switching with an eye toward neuromorphic memory applications[17]. Our goal was to find a compound with the highest optical contrast between amorphous and crystalline states in order to realize multi-level optical switching with a high signal-to-noise ratio. The composition range mapped was selected based on the lack of detailed phase distribution and optical property information near known PCM phases. We tasked CAMEO to find the composition with the largest difference in the optical bandgap $\Delta E_g$ and hence optical contrast between amorphous and crystalline states. We have discovered a naturally-forming stable epitaxial nanocomposite at a phase boundary between the distorted face-centered cubic Ge-Sb-Te structure (which we refer to as FCC-Ge-Sb-Te or simply GST) phase region and phase co-existing region of GST and Sb-Te whose optical contrast is superior to the well-known GST225 or other compounds within the Ge-Sb-Te ternary.

CAMEO satisfies many attributes of a robot scientist, as diagrammed in Figure 1. The modular algorithm has 'at its fingertips' a collection of information - knowledge of past experiments both physical and computational, materials theory, and measurement-instrumentation science. The algorithm uses this knowledge to make informed decisions on the next experiment to perform in the pursuit of optimizing a materials property and/or maximizing knowledge of a materials system. For example, at each iteration the set of possible phase maps are identified and ranked by Bayesian likelihood, given analysis results of the measured materials. Phase map and functional property likelihoods establish scientific hypotheses and drive further phase mapping and materials optimization and are also presented to the human-in-the-loop who can then (optionally) provide guidance. Specifics are presented in Methods. CAMEO controls lab-based characterization equipment in real-time to orchestrate its own experiments, update its knowledge, and continue its exploration. The more specific implementation of Eqn 1 is shown in Figure 2.

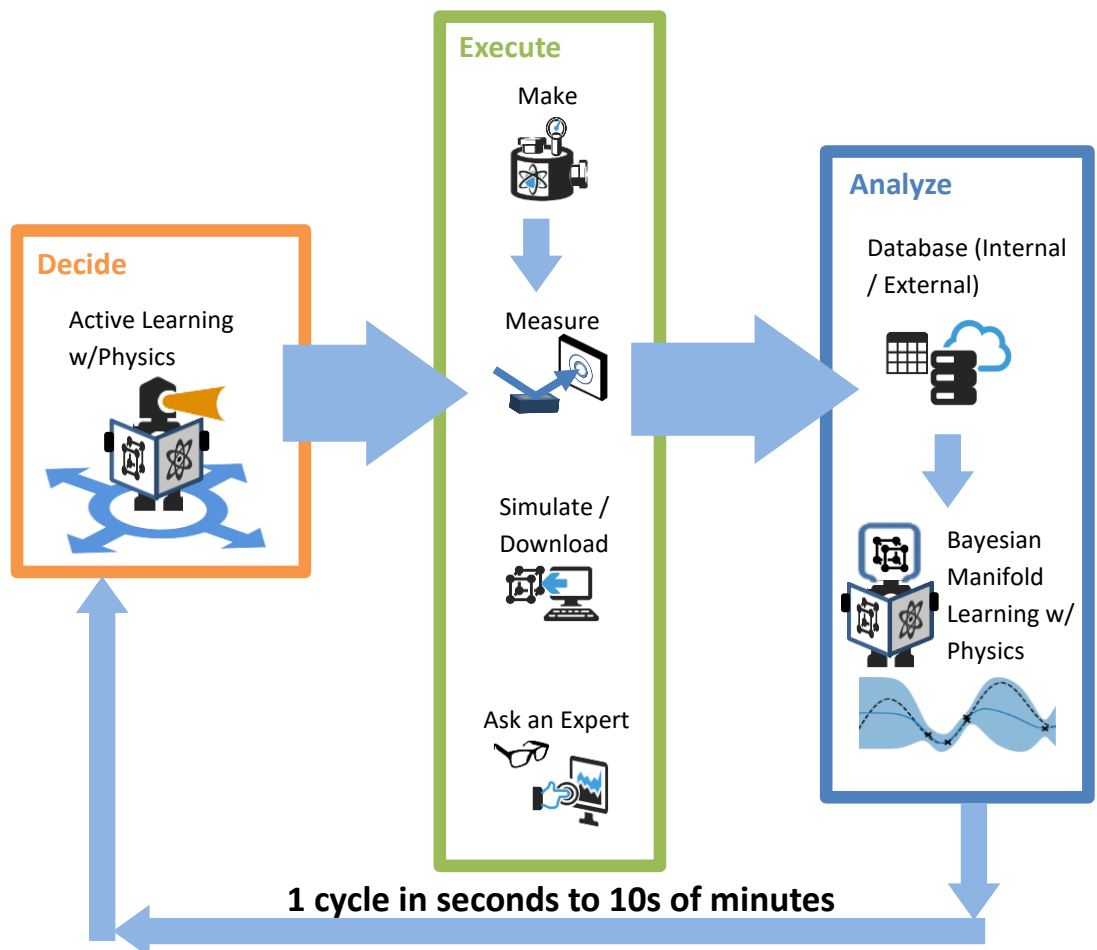

*Figure 1. Closed-loop Autonomous Materials Exploration and Optimization (CAMEO). The autonomous cycle begins with loading data from databases including composition data for the materials on the composition spread and computed materials data from the AFLOW.org[18] density functional theory database. The collected data is then used to begin analysis of the data using physics-informed Bayesian machine learning. This process extends knowledge of structure and functional property from materials with data to those without, predicting their estimated structure and functional property, along with prediction uncertainty. Physics-informed active learning is then used to identify the most informative next material to study to achieve user-defined objectives. For this work, active learning can select the next sample to characterize through autonomous control of the high-throughput X-ray diffraction system at a synchrotron beamline or it can (optionally) request specific input from the human-in-the-loop. Future implementations will include autonomous materials synthesis and simulation. The data collected from measurements and from human input are added to the database and used for the next autonomous loop. For more information, see Methods.*

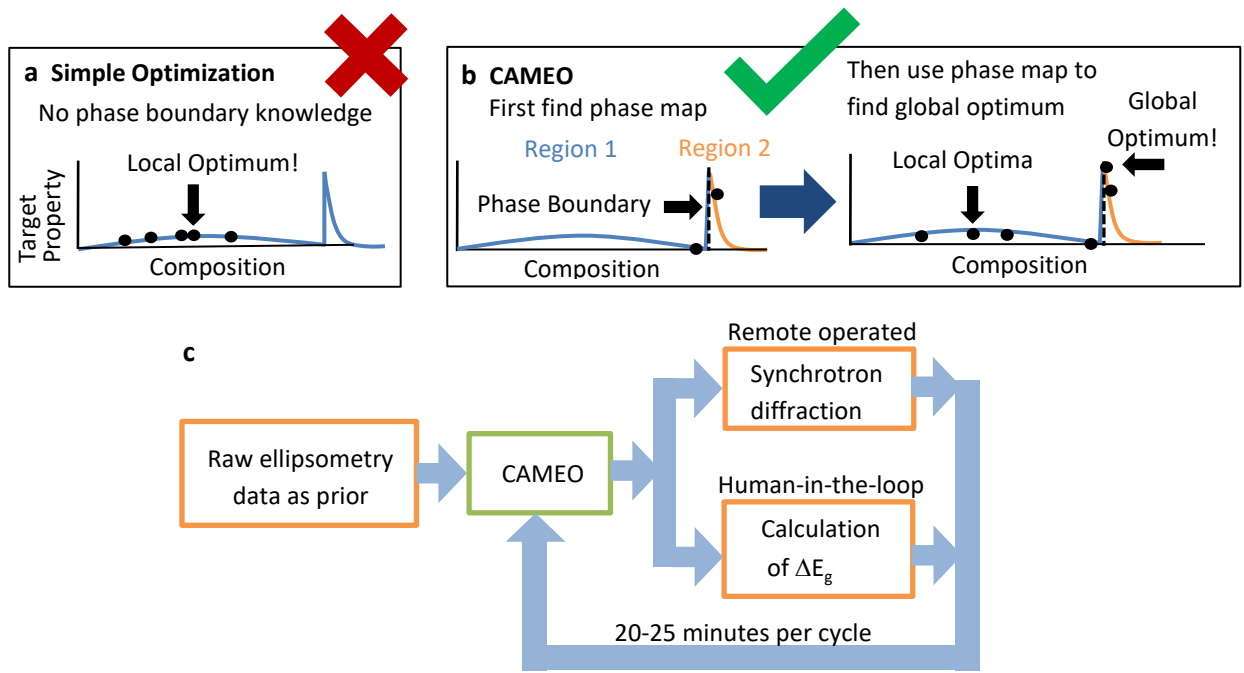

*Figure 2. Comparison of materials optimization schemes. a) Simple optimization seeks to identify the property optimum with a mixture of exploration and exploitation without knowledge of the composition-structure-property (CSP) relationship. These methods are more likely to get caught in local optima. b) The phase-map-informed optimization scheme exploits CSP relationship by recognizing that the property is dependent on phase, thus including phase mapping in the search for the optimum. i) Phase-mapping steps and ii) materials optimization step that exploits knowledge of the phase boundaries. This allows a search for phase region dependent optima. c) The Ge-Sb-Te CAMEO workflow began with incorporating raw ellipsometry data as a phase-mapping prior. On each iteration CAMEO selects a material to measure for X-ray diffraction and concurrently requests an expert to calculate $\Delta E_g$ for that material. Each cycle takes 20-25 minutes.*

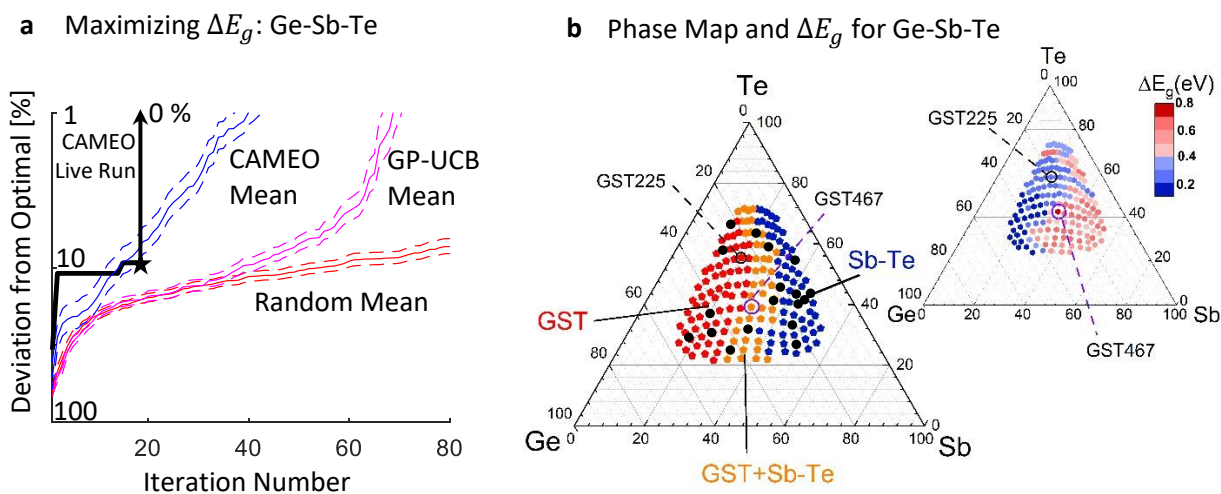

*Figure 3. Discovery of $Ge_4Sb_6Te_7$ (GST467). a) Optimization of the Ge-Sb-Te system, with the objective of identifying the material with the largest $\Delta E_g$, bandgap difference between amorphous and crystalline states. Performance is shown for: the CAMEO live run (black) with GST467 discovered on iteration 19 (black star); for the mean and 95% confidence interval of 100 CAMEO runs computed post data collection (blue), 100 runs of Gaussian Process – Upper Confidence Bounds (GP-UBC, magenta), and random sampling (red). b) Left: Structural phase map for the crystalline Ge-Sb-Te composition spread. Complete phase map constructed after all the diffraction measurements (beyond the live CAMEO run described in a)) is shown. Structural*

*phase regions are color-coded as single-phase FCC-Ge-Sb-Te (GST) structure region (red, GST), single-phase Sb-Te region (blue), and region where GST and Sb-Te phases co-exist (orange). The materials measured during the live CAMEO run are indicated in black. Right: Complete mapping of $\Delta E_g$ following analysis of entire ellipsometry data (beyond the live CAMEO run). The discovered GST467 and GST225 are indicated in both maps.*

CAMEO is based on the fundamental precept that in navigating compositional phase diagrams, enhancement in most functional properties is inherently tied to the presence of particular structural phases and/or phase boundaries. The strategy is therefore broadly applicable to a variety of topics with disparate physical properties. The method was first benchmarked and its hyperparameters tuned using a previously characterized composition spread where an entirely different physical property was optimized (See Methods). It was then successfully used to discover a new photonic PCM composition whose $\Delta E_g$ (between crystalline and amorphous states) is up to 3 times larger than that of the well-known $Ge_2Sb_2Te_5$.

For this task, scanning ellipsometry measurements were performed on the spread wafer with films in amorphous (initial) and crystalline states ahead of the CAMEO run, and the raw ellipsometric spectra data were incorporated as a phase-mapping prior (See Methods). Thus, the algorithm makes use of information regarding phase distribution across the spread that is "hidden" in the unreduced complex spectroscopic data, which vary non-trivially across the spread. At each iteration, CAMEO identifies the next material to query, indicates the material to the experimentalist (human-in-the-loop) who performs the intensive task of processing the raw optical data to extract $\Delta E_g$. In parallel, CAMEO remotely controls scanning of the synchrotron beam to collect X-ray diffraction data from the spread wafer with films in the crystalline state. CAMEO first seeks knowledge of the phase map until 80% convergence, and then switches to material optimization (See Methods section M1). This procedure identified the material with the largest $\Delta E_g$ over 19 iterations taking approximately 10 hours, compared to over 90 continuous hours for the full set of 177 composition spots. CAMEO provides an approximate maximum average 20-iteration lead over the best alternative Gaussian Process – Upper Confidence Bounds (GP-UCB) focusing on $\Delta E_g$ optimization in the composition space, and random sampling, both common benchmarks in BO.

As seen in Fig. 3(c), the optimal composition identified here lies at the boundary between the FCC-Ge-Sb-Te (GST) phase region and the region where there is co-existence of GST and Sb-Te phases. The average composition of the region is $Ge_4Sb_6Te_7$, and henceforth we refer to the region as GST467. Its $\Delta E_g$ is found to be 0.76 ± 0.03 eV, which is nearly 3 times that of GST225 (0.23 ± 0.03 eV). To investigate the origin of the enhanced $\Delta E_g$ of GST467 at the phase boundary, we have performed high-resolution transmission microscopy of this composition (Fig. 4(a)) which revealed a complex nanocomposite structure consisting of GST and Sb-Te phases. As seen in the figure, the phases have grown coherently with the relationship GST Fm-3m (111)//SbTe (001). (See Methods for details.)

This boost in $\Delta E_g$ indeed directly leads to large enhancement in optical contrast as captured in $\Delta k = k_c - k_a$, the difference in the extinction coefficient (between amorphous ($k_a$) and crystalline states ($k_c$)) extracted from the ellipsometry data at different wavelengths (Fig. 4(b)). $\Delta k$ for GST467 is 60 to 125 % larger than that of GST225 in the 1000 – 1500 nm wavelength range. The superior physical properties of GST467 shown here were reproduced on multiple composition spread wafers.

We have fabricated photonic switching PCM devices based on the discovered GST467 nanocomposite. With a sequence of laser pulses (energy and pulse width) with varying amplitude sent through the device, the material can be switched between the crystalline and amorphous phases (Fig. 4(b)). The device made of the nanocomposite GST467 thin film was found to be stable up to at least 30,000 cycles indicating the high reversibility of the crystallization and quenching processes of the coherent nanocomposite. The one-to-one comparison between the devices fabricated with our GST225 and GST467 films here (Fig. 4(d)) shows that GST467 device exhibits a much-enhanced switching contrast resulting in up to 50% more in the number of interval states, important for photonic memory and neuromorphic devices[19, 20].

Recent reports of nanostructured PCM materials including multilayer and superlattice thin films have highlighted the crucial roles interfaces and defects play in their switching mechanisms leading to faster switching speed and lower switching energies[21, 22]. Our finding of GST467 exhibiting significant boost in $\Delta E_g$, and consequently larger optical contrast, underscores the effectiveness of naturally-forming nanocomposites as another approach to enhancing performance of PCM materials, especially for optical switching devices. It is the presence of epitaxial nano-pockets of the SbTe phase in GST467 which is locally modifying the resonant bonding in the GST matrix resulting in the lowered optical bandgap in the crystalline state, which in turn leads to the larger $\Delta E_g$.

The discovery of a novel PCM material demonstrate that systems similar to CAMEO will fulfill the primary goals of materials design by accelerating the discovery and collection of materials knowledge, streamlining the experiment cycle, improving control over experimental variability, and improving reproducibility, thus *improving trust in scientific results*. They will also generate reference and benchmark datasets – automatically processed, analyzed, and converted to actionable knowledge with all associated metadata, for developing and *improving trust in machine learning tools*. Further benefits include automatic knowledge capture to maintain institutional knowledge, maximizing information gain, and reducing use of consumable resources such as expert time, freeing up experts to work on higher level challenges. Research at the synchrotron exemplifies these resource demands and limitations, where obtaining scientist and equipment time is difficult or expensive. And potentially most impactful, placing labs under the control of AI may greatly reduce the technical expertise needed to perform experiments, resulting in a greater 'democratization' of science[23]. In turn, this may facilitate a more distributed approach to science, as advocated by the materials collaboratory concept[24].

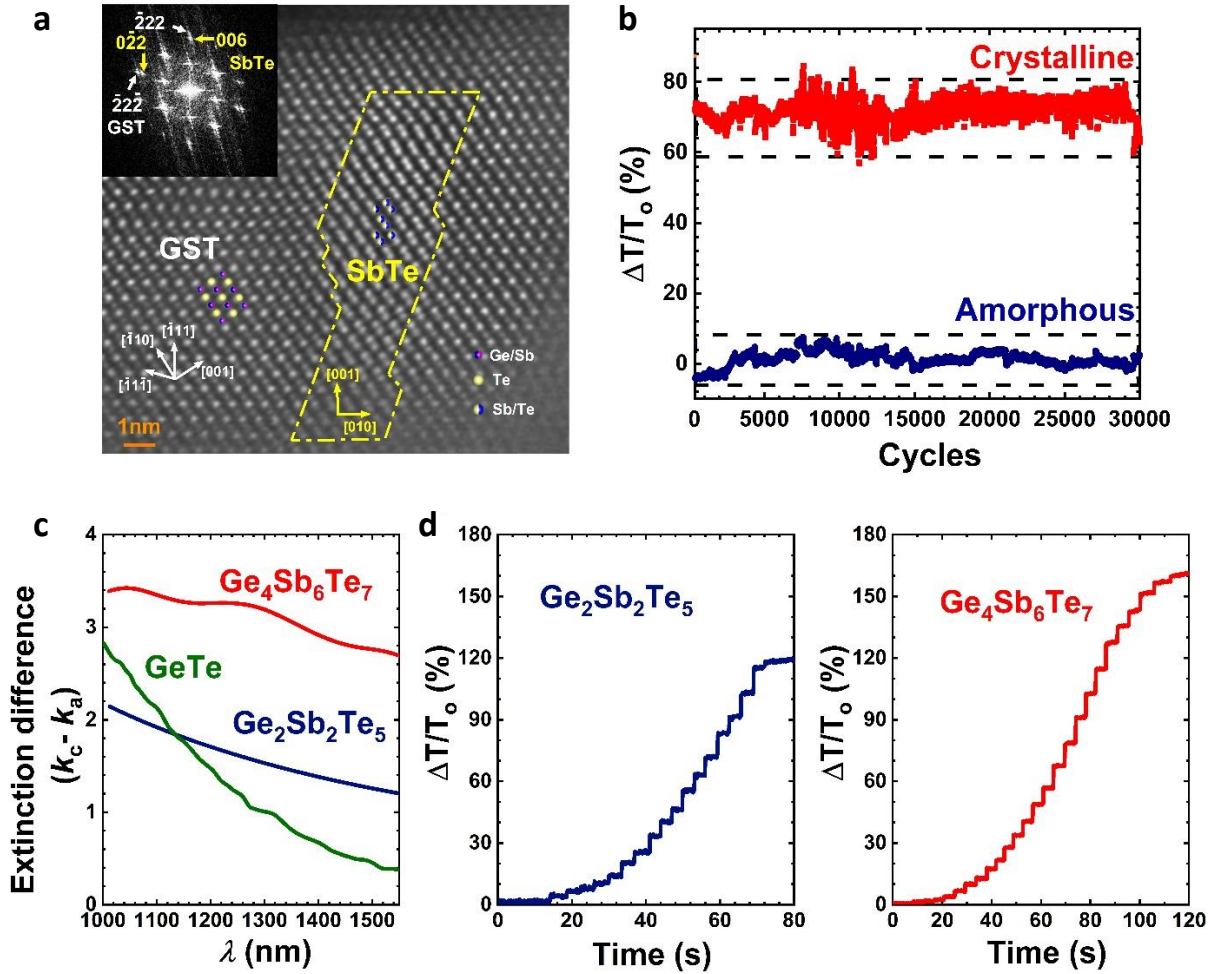

Figure 4. Nanostructure and device performance of ($Ge_4Sb_6Te_7$) GST467. Newly identified phase-change memory material GST467 shows large optical contrast ideal for photonic-switching device applications such as neuromorphic computing. (a) High resolution transmission electron microscopy image reveals formation of coherent nanocomposite of GST structure matrix and SbTe. The dotted lines denote the atomically sharp interface. The FFT (inset) of this region indicates structural similarity of the adjacent phases; (b) endurance of the GST467: it is stable over 30,000 cycles indicating the robustness of the nanocomposite structure defined by local composition variation. The dotted lines indicate the range of each state in relative optical transmission $\Delta T/T_o$ at 1500 nm. Laser pulses were 50 ns with 183 pJ for quenching and 500 ns with 3.3 nJ for crystallization. The fluctuations in $\Delta T/T_o$ are due to thermal fluctuation of the device measurement set up; (c) comparison of the optical contrast here indicated by difference in the extinction coefficient k between crystalline and amorphous phases($k_c$-$k_a$) for the wavelength range of 1000 to 1500 nm for various compositions within Ge-Sb-Te system. GST225 and GST467 data are from this work. The GeTe data are from Ref. 25. GST467 shows higher extinction difference over other known compositions; (d) one-to-one comparison of GST225 (left) and GST467 (right) for multi-level switching in optical transmission at 1500 nm ($\Delta T/T_o$) using 500 ns, 6 pJ pulses : GST467 having larger optical contrast results in substantially more states than GST225. Device fabrication and characterization details are in Methods.

## Methods

A description of the closed-loop, autonomous system for materials exploration and optimization (CAMEO) scheme can be found in M1 beginning with a detailed description of results and a description of materials and device synthesis and characterization in section M2. The description of CAMEO is broken down into the subsections: M1a Detailed results, M1b Initialization and data preprocessing, M1c Phase mapping, M1d Knowledge propagation, M1e Active learning, M1f Statistics and performance measures. The materials and device section is broken down into the subsections: M2a Sample fabrication, M2b Mapping of phase-change temperature, M2c Structural Mapping, M2d High-angle annular dark-field scanning transmission electron microscopy (HAADF-STEM) of $Ge_4Sb_6Te_7$ (GST467), M2e Modeling and calculation of the ellipsometry spectra, M2f ST467 photonic device fabrication and measurement.

**M1 CAMEO**
CAMEO's methodology follows the diagrams of Figure 1 and 2c (main text), where active learning drives measurement and expert input. Active-learning-driven synthesis and simulation are excluded for this work. The materials are pre-synthesized as a composition spread and the AFLOW.org density functional theory (DFT) simulations are run prior to CAMEO's control of the X-ray diffraction measurement. The combinatorial library is physically loaded into the high-throughput X-ray diffraction system, and any data captured from external or internal databases is automatically imported into CAMEO. All preliminary data is analyzed to build the first estimated phase map along with uncertainty quantifications. This kicks off the iterative autonomous process where the phase map and material property estimate and estimate uncertainty are used to inform the active-learning-driven selection of the material to query next. At each iteration, CAMEO selects a material to study and requests and obtains structure and functional property data for the query material, with automatic X-ray diffraction pre-processing. In parallel, results and predictions are presented to the expert user and pertinent knowledge is captured from the expert. All gathered knowledge is then stored in a database. A description of the capabilities of CAMEO are presented in Table M1.

CAMEO's specific implementation of Eqn 1 is shown in Eqn M1:

$$g(x) = \begin{cases} P(x), & c < 80\,\% \\ F(x_r) = \mu(x_r) + \beta\sigma(x_r) + \gamma d(x_r), & else \end{cases} \quad Eqn\ (M1)$$

The first set of iterations maximize phase map knowledge until the estimated phase map convergences to the user defined threshold $c$, at which point the system switches to materials property optimization. A separate Gaussian Process is fit to each individual phase region for the functional property, allowing for phase region dependent hyperparameter optimization. This exploits the CSP relationship to improve functional property prediction accuracy, accelerate materials optimization, and provide potential computational resource savings. The phase regions are then ranked by the maximum expected functional property value and the top $R$ regions are selected for optimization, with $R$ a user defined variable. Here $R$ is set to 1. Optimization balances exploitation and exploration through the mean $\mu(x_r)$ and weighted variance $\beta\sigma(x_r)$ (the iteration dependent $\beta$ follows [27] and is described below). The optimization acquisition function also allows the user to target points closer or further from phase boundaries via $\gamma d(x_r)$, where $d(x_r)$ is the distance from point $x_r$ to the nearest phase boundary and $\gamma$ is a user defined parameter – negative (positive) to emphasize points near the edge (center) of the phase region. Here the value is set to 10. Myopia to particular phase regions can be removed with an additional exploration policy.

Pre-synthesized (pseudo) ternary combinatorial spreads are used to provide a pool of hundreds of materials to investigate. While for this demonstration the autonomous system must select samples from the given pool of pre-synthesized samples, this is only a limit of the current physical experimental system and not a limit of the presented ML methodology.

**M1a CAMEO Detailed results**
CAMEO was benchmarked on a material system previously extensively studied by the authors[28]. Efficacy was compared to a range of alternative methods as shown in Fig. M1 with phase mapping performance measured with the Fowlkes-Mallows Index (FMI) and Bayesian optimization performance measured by minimum percent deviation from optimal. The mean performance and 95% confidence intervals over 100 iterations are plotted in Fig. M1. The algorithm provides significant accuracy improvement and lower variance in phase mapping. Additionally, each level of increased physical knowledge further accelerates phase mapping. The benchmark optimization challenge was to maximize a functional property that is a simple function of composition with one broad, dominant peak. For this simple challenge, CAMEO provides improved performance compared to the next best optimization scheme – Gaussian Process Upper Confidence Bounds (GP-UCB). For more information about the benchmarking process see Section 3.

Once tuned, CAMEO was placed in active control over the high-throughput X-ray diffraction system at SLAC and a commercial in-house diffraction system. Here, the material optimization goal was to identify an optimal phase change material in the Ge-Sb-Te system, characterized by maximizing $\Delta E_g$ – the difference between the amorphous and crystalline optical bandgap. Scanning ellipsometry measurements were performed on the spread wafer in amorphous and crystalline states ahead of the CAMEO run, and we fed the unprocessed ellipsometric spectra as a prior for building the phase map model. At each iteration, the query material was indicated to the experimentalist (human-in-the-loop) who then performed the intensive task of processing the raw optical data to obtain $\Delta E_g$ and provided this data to CAMEO (See section M2e for full description). This procedure identified the material with the largest $\Delta E_g$ over 19 iterations taking approximately 10 hours, compared to 90 hours for the full set of 177 materials. A post-analysis is shown in Fig. 4 (main text), where 100 runs are performed comparing CAMEO to alternative methods. CAMEO provides an approximate maximum average 30-iteration lead over GP-UCB. More importantly, the algorithm is able to mine and make use of information regarding phase distribution across the spread hidden in the complex raw spectroscopic data.

**M1b System Initialization and Data Pre-processing**
**Physical system initialization**
The system is initialized by loading the composition spread into the X-ray diffraction system, either the Bruker D8 or the SSRL diffraction synchrotron beamline endstation. For the SSRL system, a network connection is used for sending commands to the X-ray diffraction system via the SPEC interface[29]. Exposure time for each point measurement was 15 seconds.

**Importing external data: ICSD and AFLOW.org**
The user first indicates the material system of interest. A database of known stable phases, derived from past phase diagrams, is then used to automatically identify pertinent phases. Structure data is then automatically assembled for these phases from the Inorganic Crystal Structure Database (ICSD) – a database of critically evaluated experimental structures, and the AFLOW.org[16] density functional theory database. All retrieved structures are then used to generate simulated diffraction patterns through a call to Bruker's Topas*,[30]. After data is collected from the databases, the pool of material samples is updated to contain both the samples on the composition spread and those derived from databases. Previously it was shown that external structure data improved phase mapping performance in the case of exhaustive data collection[30]. For this work, the AFLOW.org computed ternary energy hull is imported and converted to region labels which are used as phase region (i.e. cluster) priors, see Fig. M2.

**Initialize phase mapping**
Phase mapping is initialized with a user-selected expected number of phase regions for the material system, 5 for Fe-Ga-Pd and 10 for Ge-Sb-Te. While this number is used to initialize the phase map model, the phase mapping technique will converge to either a larger or smaller number of phase regions as described in the GRENDEL (graph-based endmember extraction and labeling) section. All other phase mapping hyperparameters were optimized on the benchmark system, and these values were used without modification for the Ge-Sb-Te system. Other default parameters include: graph distance multiplier is 1.2 and max number of iterations is 100.

**Selection of first sample to seed processes**
If prior material structure data is imported, such as data from AFLOW.org, that knowledge is used to initialize phase mapping (see Incorporating Prior), with the active learning criterion used to select the most informative material to query next. However, if no such prior data is used, the first sample queried can be selected randomly or using some other informative process. For benchmarking, the initial material was selected at random with uniform probability. For the live application to the Ge-Sb-Te system, the first sample was selected to be the one at the composition center of the materials on the composition spread. This sample was selected as it is potentially the most informative, given no other knowledge of the samples. The live run for the Ge-Sb-Te system completed after all the materials were measured, allowing for later analysis of active learning methodologies. To compare these methods, the initial material was again selected at random with a uniform prior.

**Measurement and Data Pre-processing: Collection, integration, background subtraction**
Once the next query material has been identified, the system then measures the query material for X-ray diffraction using a programmatically generated script via SPEC for the SLAC high-throughput system or a GADDS script for the Bruker system. For the Bruker system, the diffraction image is integrated into a diffraction pattern automatically, and for the SLAC system, integration is performed[31]. The background signal is then automatically identified and subtracted.

The background signal from sample to sample can vary significantly, requiring a background subtraction method capable of handling these variations. For both the SLAC and Bruker diffraction measurements, Matlab's envelope function with the 'peak' setting and the parameter value of 50 was used to identify and remove the background curve.

**M1c Phase Mapping**
**Main method: GRENDEL – List of Physical constraints.**
Phase mapping was performed using the physics-informed phase region and constituent phase identification method described in [32]. This method represents the material composition space as a graph, where each material is represented by a vertex that is connected by edges to neighboring materials in composition space (or wafer x-y coordinate). Neighboring materials are defined by Voronoi tessellating the composition space[32]. Mathematically, $G = \{V, E\}$, where $V$ is the set of vertices, $E$ is the set of edges with all edge weights set to 1. $G$ is used to define a Markov Random Field[33] where materials identified with the same vertex label belong to the same phase region, and each phase region is described by a set of constituent phases. This method encodes a list of physical constraints through the methods listed in **Error! Reference source not found.**:

This method identifies a phase map for hundreds of samples in tens of seconds, on the same order of X-ray diffraction measurements at SSRL which typically takes 30 seconds, and measurements on the Bruker D-8* which takes over 10 minutes.

GRENDEL hyperparameters include the MRF segmentation (i.e. graph cut) weight and the balance between the material-phase region matrix based on clustering and that based on phase mixture[32]. As the graph cut weight is increased, a greater number of clusters becomes possible, increasing the phase mapping performance using the measured described in the text, while also increasing cluster complexity. For the Fe-Ga-Pd a graph cut weight of $w_{gc} = 100 * n^3/5^3$ was found to output the desired number of clusters $n$. The full set of phase mapping parameters are provided in Table M1.

During the GRENDEL process, if the number of clusters drops below 90 % of the number of clusters used when starting the process, GRENDEL is terminated and the computed phase map labels and constituent phases from the previous internal GRENDEL iteration are output.

**Comparison method: HCA**
Phase mapping performance of GRENDEL is compared to phase labels identified with Agglomerative Hierarchical Cluster Analysis, where samples of similar diffraction patterns are grouped together into clusters which are then associated with the underlying phase regions. A user defined number of phase regions (i.e. clusters) is used as well as the cosine metric and the average method for computing cluster-cluster dissimilarity.

**Phase mapping prior**
Material property data is incorporated into the MRF model as a prior through the edge weights of the composition graph $G$, where the original edge weights of $G$ are modified by a functional property graph $G_p$ with edge weights of 0 (disconnected) or 1 (connected) and $f: E, E_p \to E$. If $e \in E \cap E_p$ then $e = 1 + \Delta_e$ else $e = 1 - \Delta_e$. The value of $\Delta_e$ was varied for the benchmark material system and the value of $\Delta_e = 0.5$ selected as it demonstrates clear improved phase mapping performance during the first active learning selected measurements and worse performance near the end of the run. This is to be expected as prior knowledge can benefit initial analysis but can overwhelm knowledge gained from data if the prior is weighted too heavily. A smaller (larger) value of $\Delta_e$ demonstrates a smaller (larger), similar effect.

For the benchmark system an AFLOW.org based phase map prior was used, as shown in Figure M2, where the AFLOW.org tie-lines are used to define regions. Points that fall in the same region are given the same label, resulting in a prior for phase mapping. For materials that share a graph edge and a clustering label, the edge weight in $E_p$ is set to 1, otherwise the edge connecting them is removed from $E_p$.

For the Ge-Sb-Te material system, the prior was determined based on optical data collected. For each material, the complex reflectance ratio amplitude $\psi$ and phase difference $\Delta$ for the amorphous and crystalline phases were collected for the set of angles $\theta = \{50°, 55°, 60°, 65°, 70°\}$ relative to the laser's plane of incidence, creating 20 spectral measurements for each material consisting of different measurement types $m \in \{\psi^{crystalline}, \psi^{amorphous}, \Delta^{crystalline}, \Delta^{amorphous}\}$. Example optical data used for the prior is shown in Fig. M3.

To define a prior for the phase diagram, the set of all spectra are reduced into a set of similarity weights defining a similarity of 0 or 1 for each pair of samples, which can then be used to evaluate $E_p$. The following equations are used for mapping of spectra to similarity values. First the Euclidean difference between each set of materials $(i, j)$ is computed for each spectral measurement type and angle $\{m, \theta\}$. These differences are then averaged for each pair of samples $(i, j)$ over the set of angles $\theta$ and then normalized to between 0 and 1 for each measurement type $m$. These values are then averaged again over measurements $m$, resulting in a final dissimilarity value for each pair $(i, j)$. A threshold is then used to convert the continuous dissimilarity values to 0 or 1, defining whether an edge between $(i, j)$ exists (1) or does not (0). The threshold of $D_{Threshold} = 0.07$ was selected to achieve a ratio of $|E_p|/|E| = 0.49 \approx 0.5$, i.e. the prior removes approximately half the edges from the initial graph.

$$D_{i,j}^m = \underset{\theta}{\text{mean}}[d_{Euclidean}(m_{i,\theta}, m_{j,\theta})]$$

$$D'^m_{i,j} = [D_{i,j}^m - \min D_{i,j}^m] / [\max D_{i,j}^m - \min D_{i,j}^m]$$

$$\bar{D}_{i,j} = \underset{m}{\text{mean}}[D'^m_{i,j}]$$

$$G_p = \bar{D}_{i,j} < D_{Threshold}$$

**M1d Knowledge Propagation**
**Phase Mapping Knowledge Propagation**
Once the phase map has been identified for the given data, the phase region labels must be propagated to the materials that have yet to be measured for structure. To exploit the graph-described data manifold, the semi-supervised learning technique Gaussian random field harmonic energy minimization[35] (HEM) is used. HEM computes the likelihood of each material belonging to each phase region and then assigns each material to the phase region with the greatest likelihood, thus defining the most likely phase map for the full set of materials on the composition spread. The edge weights $E_p$ define the similarity matrix used to define the graph

Laplacian.

**Phase Mapping Knowledge Propagation - Comparison method: Nearest Neighbor (NN)**
The phase mapping knowledge propagation harmonic energy minimization method is compared to the use of 1-nearest neighbor, where any material without a phase region label takes on the label of its 1st nearest neighbor with a label. First nearest neighbor was implemented using MATLAB's knnsearch function with default parameters.

**Functional property Knowledge Propagation: Gaussian Process Regression**
GPR was implemented using MATLAB's 'fitrgp' function with default parameters.

**M1e Active Learning**
In the Bayesian optimization[36] literature, the following formalism is used:

$$y = f(x) + \epsilon$$

$$x_* = \mathrm{argmax}_x(f(x))$$

where $y$ is the target property to be maximized, $x \in \mathbb{R}^d$ is the set of material synthesis and processing parameters to be searched over, $f(x)$ is the function to optimize, $\epsilon$ is typically independent stochastic measurement noise, and $x_*$ defines the material synthesis and processing parameters that result in the maximal material property (for the minimum, replace argmax with argmin). When $f(x)$ is unknown, a surrogate model is used to approximate it based on given data. The surrogate function is then used to identify the best next material to study. Each subsequent material is selected to identify the optimal material $x_*$ in the smallest number of experiments possible. Identifying extrema of a function involves a balance between *exploiting* prior data to identify nearby extrema and *exploring* to identify extrema far from prior data. An alternative active learning objective is to select experiments that will best improve the overall prediction accuracy of the surrogate model, or in other words, select experiments to most efficiently learn the unknown function $f(x)$. Such a campaign learns the general trends of $f(x)$, which is highly useful when attempting to quantify anomalous behavior of novel materials.

**Active Learning for Phase Mapping: Risk minimization**
The active learning method used to select the next material to query for phase mapping is based on risk minimization[35]. HEM propagates phase region labels to unmeasured material and identifies the likelihood of each material belonging to each phase region. These likelihoods can be aggregated to define the set of potential phase diagrams and their associated likelihoods. The set of potential phase diagrams form a hypothesis space of phase diagrams. Risk minimization seeks to identify the optimal material to query next for its structure that will most rapidly whittle down the hypothesis set and most rapidly hone-in on the optimal phase map for the full set of materials, i.e. minimize expected total phase region label misclassification error and equivalently maximize knowledge of the phase map.

**Active Learning Comparison Methods: Random, Sequential, and 10 % Sampling**
The risk minimization method is compared to 1) random sampling – selecting each subsequent material at random from the wafer, with a uniform prior, 2) sequential sampling – where each sample is selected in the order it appears on the wafer, and 3) where 10 % of the materials are selected in a pre-determined design. Random sampling is expected to provide increasingly poor performance relative to active learning as the search space increases in dimension due to the curse of dimensionality[37]. The pre-determined 10 % selection of materials in (3) are chosen to provide maximal coverage of the composition space. However, the use of 10 % is not a generalizable benchmark. For a given number of data points, the density of data points decreases as the dimensionality of the composition space increases, with each point describing a larger volume. The optimal number of benchmark materials is thus dependent on the expected size of phase regions. If smaller phase regions are expected, a larger number of materials will be required to identify the phase regions.

The Fe-Ga-Pd composition spread contains 278 samples. For the 10 % sampling, the 28 samples are indicated in Fig. M4(a). They were selected to provide uniform coverage of the composition space described by the spread. For the sequential sampling, the order of samples is shown in Fig. M4(b).

**Active Learning - Materials Optimization: Gaussian Process Upper Confidence Bounds**
For CAMEO and GP-UCB the iteration dependent weight parameter $\beta$ is used[27].

$$\beta = 2\log(DI^2\pi^2/6\lambda)$$

Where $D$ is the total number of samples, $I$ is the current iteration number, and $\lambda = 0.1$.

**Active Learning – CAMEO: Phase Mapping Convergence**
The phase maps identified at each iteration $i$ is compared to the iteration $(i-4)$ using the FMI performance measure. Convergence is defined as FMI >= 80 %.

**Active Learning – Materials Optimization: Benchmark System**
The target optimization for the benchmark system is maximizing remnant magnetization. One modification was made to the remnant magnetization signal: The signal saturates over a large range of the composition spread. For BO benchmarking, it is preferred that one material is identified as the optimum. As such, the saturated values were modified with a squared exponential function, in effect "hallucinating" the remnant magnetization values as if sensor saturation had not occurred, converting the signal from Fig. M5(a) to Fig. M5(b). The squared exponential function used to modify the voltage was defined in cartesian space. For the ternary composition $(a_{Fe}, b_{Ga}, c_{Pd})$:

$$x = (b/100) + (c/100) * \sin(30°), \qquad y = (c/100) * \sin(60°)$$

$$N_{xy}(\mu = (0.19, 0.05), \sigma^2 = 0.001)$$

Mapping to ternary space gives $\mu = Fe_{78}Ga_{16}Pd_{6}$.

**M1f Statistics and Performance Metrics**
**Confidence Interval**
The 95 % confidence interval was computed for the variable of interest over 100 experiments at the given iteration with:

$$\left(\frac{\sigma}{\sqrt{n}}\right) F^{-1}(p, v)$$

Where $F^{-1}$ is the inverse of the Student's t cumulative distribution function, $\sigma$ is the standard deviation, $n = 100$ is the number of experiments, $p = \{2.5\%, 97.5\%\}$, and $v = 99$ is the degrees of freedom.

**Phase mapping**
Phase mapping performance is evaluated by comparing phase region labels determined by experts with those estimated by CAMEO for the entire phase map (after the knowledge propagation step). To evaluate system performance, the Fowlkes-Mallows Index (FMI) is used, which compares two sets of cluster labels. The equations are presented below for the expert labels $l \in L$ and the ML estimated labels $\hat{l} \in \hat{L}$, where the labels are enumerated $L \to \mathbb{N}$ and $\hat{L} \to \mathbb{N}$.

If the number of phase regions is taken to be too large by either the user or the ML algorithm while the phase mapping is correct, some phase regions will be segmented into sub-regions with the dominant phase boundaries preserved. For example, peak shifting can induce phase region segmentation[30]. To ensure that the performance measures ignore such sub-region segmentation, each estimated phase region is assigned to the expert labeled phase region that shares the greatest number of samples. The number of phase regions is monitored to ensure that increases in model accuracy are not driven by increases in model complexity.

Fowlkes-Mallows Index: $FMI = TP/\sqrt{(TP + FP)(TP + FN)}$

TP $\quad \frac{1}{2}\sum_i \sum_j (l_i = l_j \,\&\, \hat{l}_i = \hat{l}_j)$

FP $\quad \frac{1}{2}\sum_i \sum_j (l_i \neq l_j \,\&\, \hat{l}_i = \hat{l}_j)$

$$\text{FN} \quad \frac{1}{2}\sum_i\sum_j (l_i = l_j \,\&\, \hat{l}_i \neq \hat{l}_j)$$

$$\text{TN} \quad \frac{1}{2}\sum_i\sum_j (l_i \neq l_j \,\&\, \hat{l}_i \neq \hat{l}_j)$$

**Bayesian Optimization**
Bayesian optimization performance is measured with minimum percent deviation from optimal, related to simple regret.

$$\text{Minimum percent deviation from optimal} = 100\,\% * \min_i(p_{target} - p_i)/p_{target}$$

$$\text{Simple Regret} = \min_i(p_{target} - p_i)$$

**M2 Materials Synthesis and Characterization**

**M2a Sample fabrication**

Amorphous thin-film composition spreads encompassing a region of the Ge-Sb-Te ternary (separated into 177 samples using a gridded physical shadow mask) were fabricated on 3-inch silicon wafers with $SiO_2$ layers (2 µm) by co-sputtering Ge, Sb, and Te targets at room temperature. Different (average) thickness composition spreads (covering the same composition range) were fabricated for different measurements: they were 20 nm, 100 nm, 200 nm, and 500 nm for optical, structural, resistance, and composition mapping, respectively. To obtain a crystalline state, some of the wafers were annealed at 300 ºC for 10 min following their characterization in the amorphous state.

The composition mapping of the spreads is measured using the wavelength dispersion spectroscopy. For every separated sample region on a spread, three random spots are measured, and the average composition value is used for the actual stoichiometry mapping in Fig. M6.

**M2b Mapping of phase-change temperature**

Upon increasing the temperature, a phase-change memory material undergoes a structural transition from amorphous to crystalline states with up to four orders of magnitude in the change of resistance. The temperature at which the resistance drop takes place can be taken as the phase-change temperature, $T_{cp}$ (Fig. M6). The entire spreads were measured in a scanning four-probe station combined with a Keithley 2400 from room temperature up to 300 ºC. $T_{CP}$ of GST467 was found to be approximately 200 ºC, which is much higher than that of GST225 (≈140 ºC). The higher $T_{CP}$ of GST467 indicates higher stability of the amorphous state of GST467 compared to GST225.

**M2c Structural Mapping**

Synchrotron diffraction on crystallized spreads was carried out at Beamline 10-2 at SLAC. In addition to the remote-controlled CAMEO run, we have also carried out diffraction of entire spreads in order to obtain the complete structural phase mapping of the probed Ge-Sb-Te region and to verify the accuracy of the phase diagram determined by CAMEO. Fig. M7 shows an example set of diffraction patterns taken across the spread. Along the marked line in the composition map, the evolution of diffraction patterns indicates phases going from the distorted FCC-Ge-Sb-Te (GST) structure region to the phase co-existence region (GST and Sb-Te) to the Sb-Te region.

SbTe ($R\bar{3}m$), Sb$_2$Te$_2$ ($P\bar{3}m$), and Sb$_2$Te$_3$ ($R\bar{3}m$) all have very similar diffraction patterns and atomic projections of the [100] zone-axis, except for different lattice periods along the [001] direction. These three phases are present across the Sb-Te region depending on the local composition on our spread. The predominant Sb-Te phase in GST467 is SbTe (below).

**M2d HAADF-STEM of GST467**
We have performed cross-sectional High-angle annular dark-field scanning transmission electron microscopy (HAADF-STEM) measurements on the GST467 thin film and found that there are nanometer-sized SbTe regions grown coherently inside the distorted cubic GST matrix as shown in Fig. 4(a) of the main text. To distinguish between similar phases (SbTe, Sb$_2$Te$_2$, and Sb$_2$Te$_3$), analysis of electron diffraction rings was carried out (not shown here), and Sb-Te phase in the GST467 was identified to be Sb$_1$Te$_1$.

**M2e Modeling and calculation of the ellipsometry spectra**
The experimental ellipsometry data (J. A. Woollam company) of the combinatorial Gs-Sb-Te spread was analyzed in the range from 200 to 1000 nm using the CompleteEASE software. The dielectric function $\varepsilon(\omega)$ used in the model contains[38] (1) a constant, (2) a Drude-type contribution for free carriers in the case of crystalline state, and (3) a Tauc-Lorentz oscillator to describe the onset of optical transition:

amorphous state: $\varepsilon(\omega) = \varepsilon_{const} + \varepsilon_{Tauc\text{-}Lorentz}(\omega)$,

crystalline state: $\varepsilon(\omega) = \varepsilon_{const} + \varepsilon_{Drude}(\omega) + \varepsilon_{Tauc\text{-}Lorentz}(\omega)$.

For the Drude model:

$$\varepsilon_{Drude}(\omega) = \varepsilon_1(\omega) + i \cdot \varepsilon_2(\omega) = 1\left(\varepsilon(\infty)\right) - \frac{\omega_p^2}{\omega^2 + \Gamma^2} + i \cdot \frac{\omega_p^2 \cdot \Gamma}{\omega \cdot (\omega^2 + \Gamma^2)}$$

where $\omega_p = \sqrt{\frac{N \cdot e^2}{m \cdot \varepsilon_0}}$, and $\omega_p$ is the plasma frequency, $\Gamma$ is the collision frequency.

For the Tauc-Lorentz model:

$$\varepsilon_{Tauc-Lorentz}(\omega) = \varepsilon_1(\omega) + i \cdot \varepsilon_2(\omega)$$

$$= \varepsilon_1(\infty) + \frac{2}{\pi} P \int_{E_g}^{\infty} \frac{\xi \varepsilon_2(E)}{\xi^2 - E^2} d\xi + i \cdot \frac{AE_0 \Gamma (E - E_g)^2}{E[(E^2 - E_o^2)^2 + \Gamma^2 E^2]} \Theta(E - E_g)$$

Where $A$ is the prefactor, $E_o$ is the peak in the joint density of the state, $E_g$ is the optical bandgap, and $\Gamma$ is the broadening parameter.

The optical parameters, e.g. refractive index $n$ and extinction coefficient $k$, as well as the optical bandgap, can be extracted from these models. In order to confirm the accuracy of the optical parameters, one needs to check if the fitting curves as well as a set of fitting parameters, e.g. thickness, carrier density, and surface roughness, can be used to analyze the experimental data of the samples for the entire spread wafer. Typically, a fitting procedure requires repeated steps in order to fine-tune the parameters manually to optimize the results, and some samples require more manual fitting steps for setting the range and the starting values of the parameters than others.

For the CAMEO run, the unprocessed raw ellipsometry data taken at each composition spot (for crystalline and amorphous states) are used as the prior. Once a spot is identified as a possible composition with enhanced $\Delta E_g$ (the difference in the optical bandgap between the amorphous and crystalline state), the fitting procedure above is carried out on the raw data, and the value of $\Delta E_g$ is computed, the process of which includes manual inspection of fitting parameters. Depending on the number of repeated steps, each computation can take up to 20 min. at a composition spot, and the $\Delta E_g$ value is then fed back to the CAMEO algorithm.

The complete mapping of the optical bandgap of amorphous and crystalline states measured and calculated from one spread are shown in Fig. M8. In the amorphous state, the Ge-Sb-Te based compounds are effectively covalently-bonded semiconductors with large optical bandgaps[39, 40]. With changing composition, there is variation in bonding leading to slight shift in the optical bandgap shown in Fig. M8. In the crystalline state, the resonantly-bonded $p$ orbitals can delocalize the carriers resulting in the reduced bandgap[38, 41–43], leading to the large contrast between the amorphous and crystalline states. In the distorted cubic phase (i.e. GST), with changing composition, the local distortion (i.e. Peierls distortion) due to vacancies[25, 41, 44] would modify the resonant bonding shifting the optical bandgap. In the Sb-Te phase, the optical bandgap also varies with the changing composition in Fig. M8. When the epitaxial nanocomposite with the SbTe phase are coherently and homogeneously grown in the GST matrix as shown in Fig. 4(a), the nano SbTe phase can act as the impurity dopant phase in the GST matrix.

**M2f Ge$_4$Sb$_6$Te$_7$ photonic device fabrication and measurement**

Photonic switching devices were fabricated out of GST467 films (Fig. M9). The 30 nm thick nanocomposite GST467 thin film was sputtered on a 330 nm thick $Si_3N_4$ layer on an oxidized silicon wafer. A 10 nm thick $SiO_2$ protection layer was then coated on the top of the GST467 thin film. Using e-beam lithography and inductively coupled plasma etching, a 1.2 µm wide photonic waveguide was fabricated. Then the GST467 thin film was patterned into disk shaped features 500 nm in diameter on the top of the waveguide, and they were encapsulated with a 200 nm thick $Al_2O_3$ layer as shown in the inset of Fig. M9.

A symmetric multi-level switching of the photonic device was investigated as shown in Fig. M9. In order to provide and precisely control the pump pulses to quench or anneal the GST467 thin film in steps, pulses from a CW pump laser were first modulated by an electro-optic modulator and then sent into an erbium-doped fiber amplifier followed by a variable optical attenuator. The output of the optical signal was collected with a photodetector. During the annealing process, a sequence of pump pulse (50 ns, 2 mW) train was applied to the photonic device. In the quenching process, a sequence of 50 ns pump pulses with gradually increased amplitude was sent into the waveguide.

NIST disclaimer

* Certain commercial equipment, instruments, or materials are identified in this report in order to specify the experimental procedure adequately. Such identification is not intended to imply recommendation or endorsement by the National Institute of Standards and Technology, nor is it intended to imply that the materials or equipment identified are necessarily the best available for the purpose.

**Acknowledgments:** We acknowledge Xiaohang Zhang for assistance with thin film characterization and Fang Ren and Doug Van Campen for assistance in diffraction experiments at SLAC. We also acknowledge Valentin Stanev, Gaurav Modi, and Daniel Samarov for discussions. **Funding**: This work was supported by ONR MURI N00014-17-1-2661, ONR N00014-13-1-0635, and ONR N00014-15-2-222. The work at the University of Maryland was partially supported by the Center for Innovative Materials for Accelerated Compute Technologies (IMPACT), one of the centers in nCORE, a Semiconductor Research Corporation (SRC) program. Diffraction performed at SLAC is supported by the U.S. Department of Energy, Office of Science, Office of Basic Energy Sciences under Contract No. DE-AC02-76SF00515. H.Z. and L.A.B. acknowledge support from the US Department of Commerce, NIST under financial assistance award 70NANB17H249. A.V.D. acknowledges support from Materials Genome Initiative funding allocated to NIST. **Author contributions**: A.G.K. and I.T. initiated and supervised the research. A.G.K. developed and implemented CAMEO algorithm with guidance from I.T., B.D. helped refine the phase mapping prior algorithm. H.Y. fabricated the Ge-Sb-Te composition spreads and thin films and coordinated their characterization. A.M. and S.S. developed and carried out the high-throughput synchrotron diffraction. C.W. and M.L. characterized the optical properties of Ge-Sb-Te composition spreads and fabricated and performed measurements on photonic devices. H.Z. conducted the transmission electron microscopy (TEM), and A.D. and L.B. helped analyze the TEM results. J.H.S. performed some of the ellipsometry measurements. R.A. discussed the bandgap behavior of GST. C.O., C.T., and S.C. developed the AFLOW framework which was used for some of the phase diagram analysis. A.G.K., I.T., and H.Y. wrote the paper with substantial input from other authors. All authors contributed to the discussion of the results. **Competing interests**: Authors declare no competing financial interest. **Data and materials availability**: Data are available in the manuscript. Other data that support other findings of this study are available from the corresponding author upon reasonable request.

# Extended Data

*Table M1. Knowledge and Control implemented in CAMEO.*

| Knowledge and Control | CAMEO |
|---|---|
| Knowledge: Past experiments both physical and computational | Automated access to experimental and density functional theory materials structure databases. Includes Inorganic Crystal Structure Database and AFLOW.org. |
| Knowledge: Materials physics theory | Phase mapping and structure theory including Gibbs phase rule via constraint programming |
| Knowledge: Materials synthesis and processing | NA |
| Knowledge: Measurement science | X-ray diffraction simulation capability using structure data as input |
| Control: Synthesis control | NA |
| Control: Characterization | X-ray diffraction: high-throughput X-ray diffraction system[26] at the Stanford Synchrotron Radiation Lightsource (SSRL) and Bruker D-8* |
| Control: Communication | GUI for user interface; Interface to databases to store and share knowledge with experts and other AIs; Network interface for instrument control |

*Table M2. List of physical constraints in [(32)] method and associated encoding methods.*

| Physical Constraint | Encoding Method |
|---|---|
| Phase regions are cohesive and phase boundaries are continuous | 1. If two or more set of vertices share the same phase region label but are not connected by vertex neighbors, differing labels are assigned to the disconnected sets. 2. The Markov Random Field smoothness constraint |
| Materials of similar synthesis and processing parameters have similar properties | 1. Markov Random Field smoothness constraint 2. Harmonic Energy Minimization for label propagation |
| Abundances of phases is non-negative | Karush–Kuhn–Tucker conditions[34] |
| X-ray diffraction intensity is non-negative | Karush–Kuhn–Tucker conditions[34] |
| Soft Gibbs Phase Rule - Upper bound limit on number of constituent phases | Upper limit on number of endmember limits allowed in each phase region |
| Identified endmembers should be physically realizable | Volume constraint on identified / predicted endmembers |

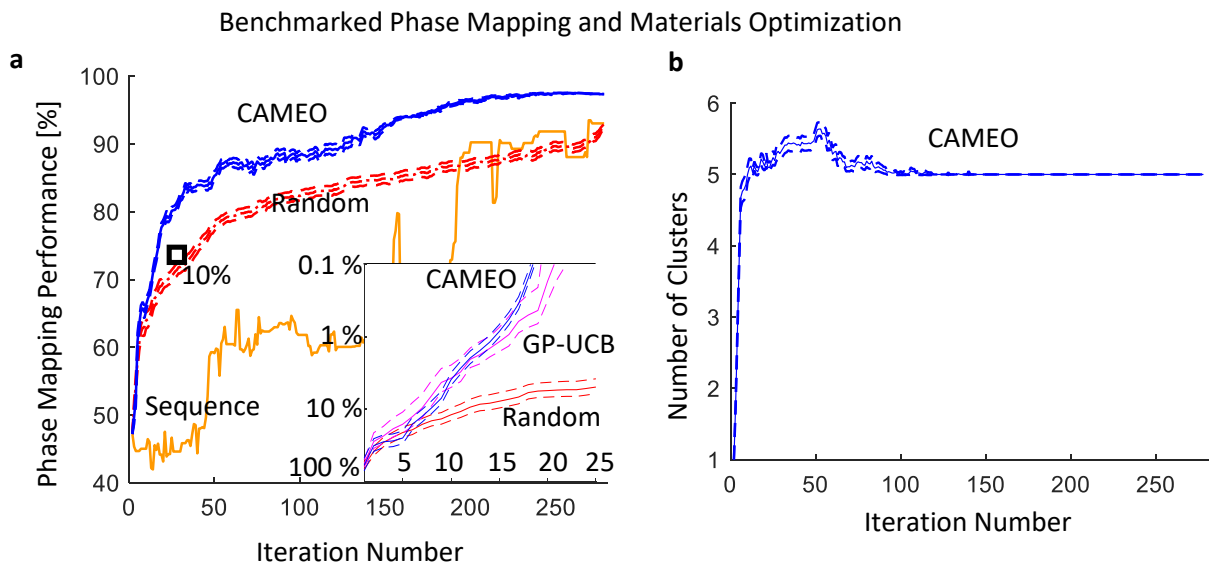

*Figure M1. Benchmarking performance. a) Main figure: Phase mapping performance demonstrating that CAMEO provides a significant advantage over the three alternatives: random sampling, sequential sampling, and measuring 10% of the samples well distributed over the composition space. Subset figure: Material optimization performance. The benchmark materials optimization challenge is highly simple with a very prominent, broad peak – a challenge that Bayesian optimization schemes excel at. Nevertheless, CAMEO provides improved results over the next best alternative, GP-UCB. Of note is CAMEO's initial lag in performance due to its initial goal of maximizing phase mapping performance. Once phase mapping performance converges, it then switches to materials optimization and shows faster performance than GP-UCB. b) The number of clusters for the benchmark dataset was initialized to 5 and while this number on average increased during CAMEO's phase mapping, it converged to 5. Demonstrating that improved performance was not due to increased complexity defined by a larger number of clusters.*

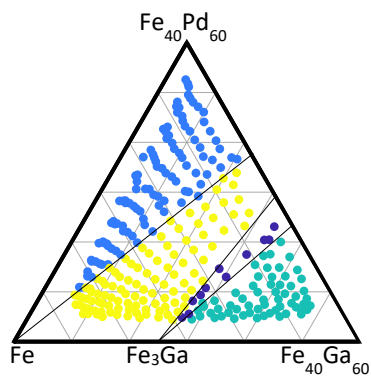

*Figure M2. Color coded phase map prior derived from AFLOW.org computed tie-lines (black lines) for the benchmark Fe-Ga-Pd material system.*

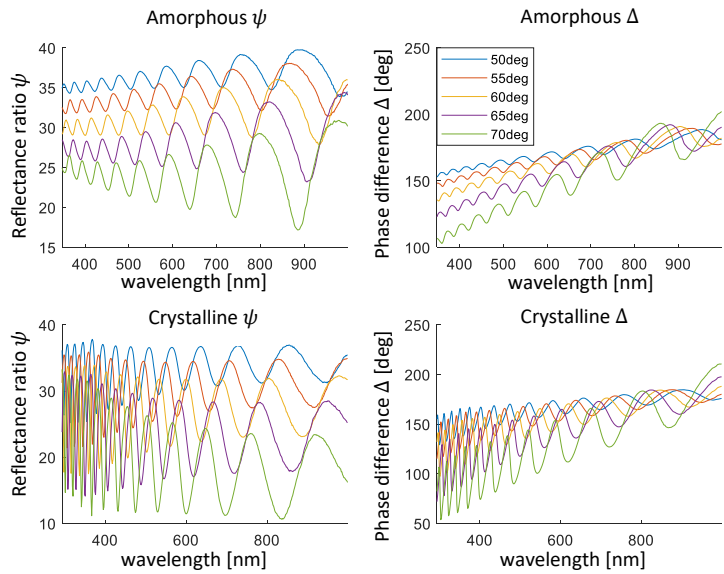

*Figure M3. Example Ge-Sb-Te optical data used for phase mapping prior.*

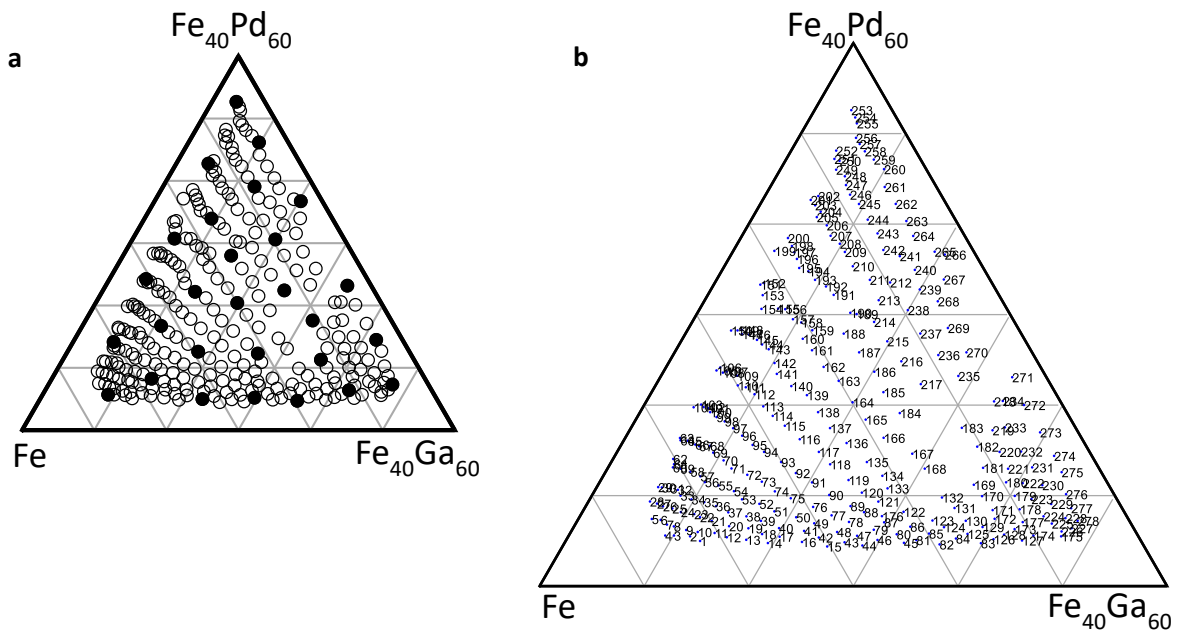

*Figure M4. a) For the 10 % material selection out of the 278 materials in the composition spread, the selected 28 materials are indicated with black filled circles. b) The order of materials measured during sequential measurement.*

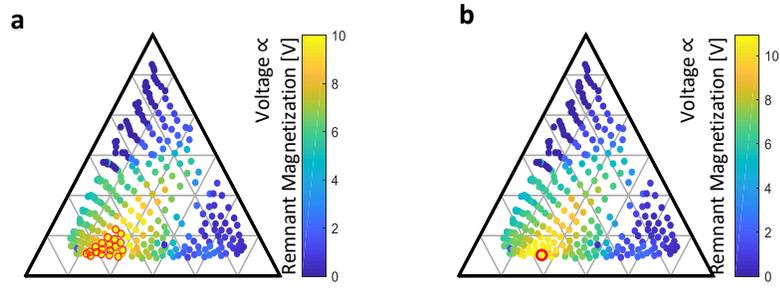

*Figure M5. Modifications made to the Fe-Ga-Pd remnant magnetization voltage signal. a) Red circles indicate the samples with saturated voltage of 10 V, b) Modified voltage by enhancing main voltage peak at $\mu = Fe_{78}Ga_{16}Pd_6$ and the maximum indicated with a red circle.*

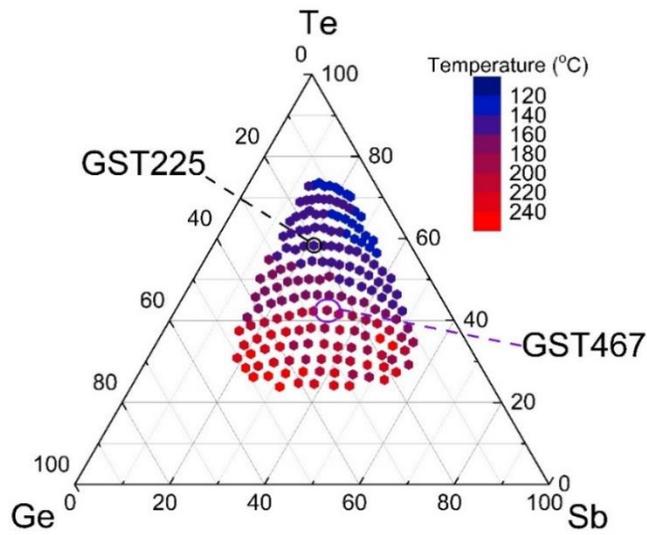

*Figure M6. Phase-change temperature mapping of the combinatorial Ge-Sb-Te spread.*

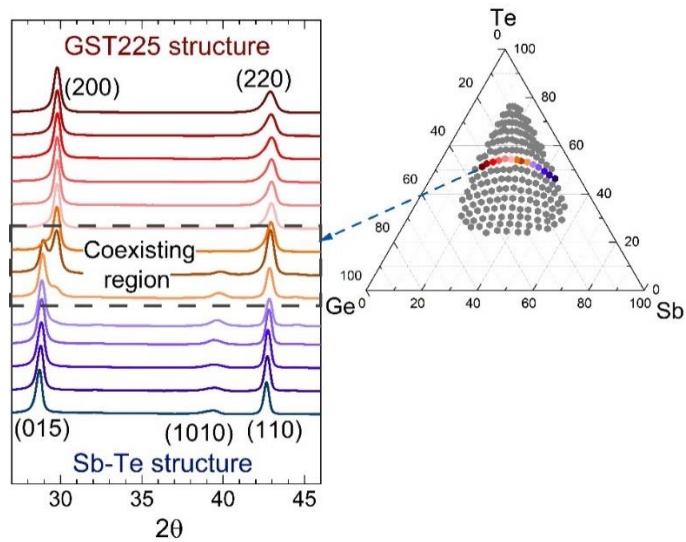

*Figure M7. Structural evolution FCC-Ge-Sb-Te (GST) structure (top) to the Sb-Te structure (bottom) across the line of composition marked in the phase diagram on the right. Peak indices are denoted.*

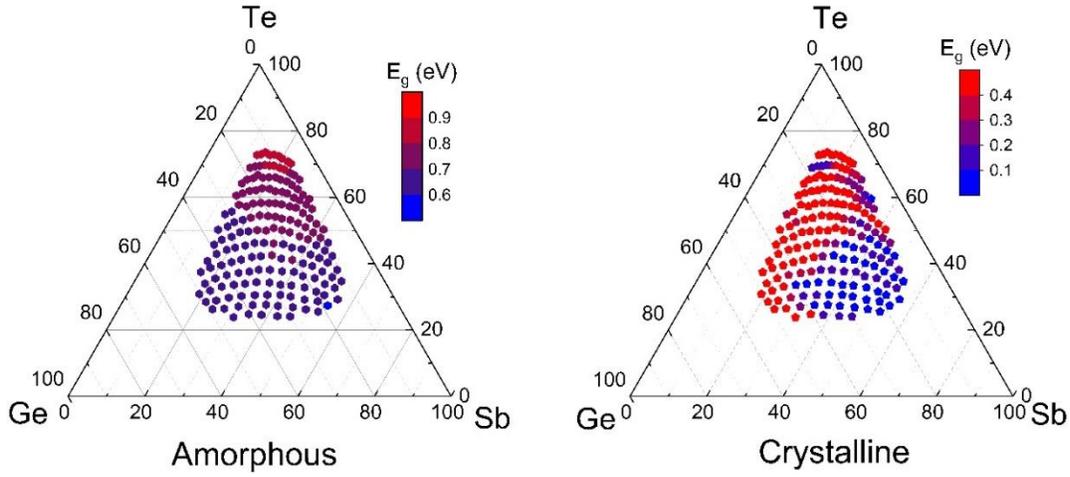

*Figure M8. The optical bandgap of amorphous (left) and crystalline (right) states for a combinatorial Ge-Sb-Te spread.*

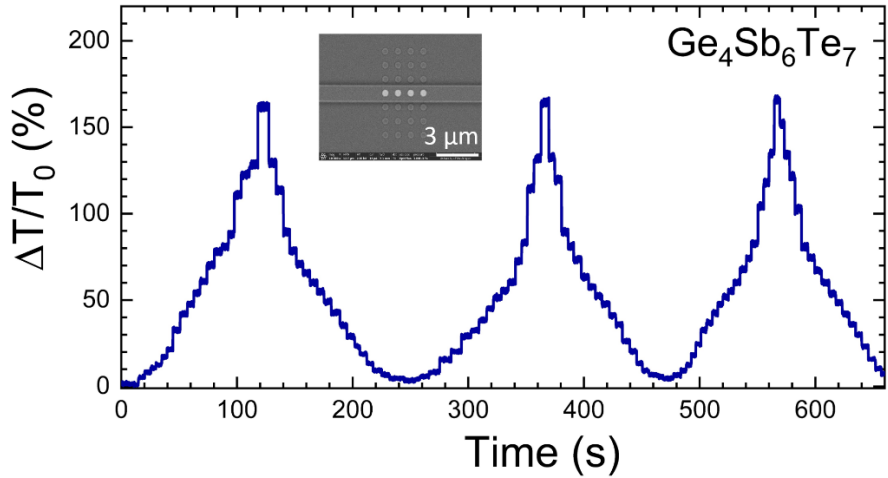

*Figure M9. The performance of the photonic device fabricated by the new nanocomposite PCM, GST467. The symmetric multi-level switching is realized. The inset is the top view of the photonic device used for multi-level switching, endurance test and comparison between $Ge_4Sb_6Te_7$ and $Ge_2Sb_2Te_5$.*